\documentclass[aps,prl,twocolumn,groupedaddress]{revtex4-1}
\usepackage{dsfont}
\usepackage{color}

\usepackage{graphicx}
\usepackage{dcolumn}
\usepackage{bm}
\usepackage{amsmath}
\usepackage{bbm}
\usepackage{placeins}

\newcommand{\be}{\begin{eqnarray}}
\newcommand{\ee}{\end{eqnarray}}

\newcommand{\ket}[1]{\left| #1 \right\rangle}               

\begin{document}

\title{Protecting conditional quantum gates by robust dynamical decoupling}

\author{Ch. Piltz}
\author{B. Scharfenberger}
\author{A. Khromova}
\author{A. F. Var\'{o}n}
\author{Ch. Wunderlich}
\email{wunderlich@physik.uni-siegen.de}

\affiliation{Department Physik, Naturwissenschaftlich-Technische
Fakult\"at, Universit\"{a}t Siegen, 57068 Siegen, Germany}


\begin{abstract}

Dephasing -- phase randomization of a quantum superposition state -- is a major obstacle for the realization of high fidelity quantum logic operations. Here, we implement a two-qubit Controlled-NOT gate using dynamical decoupling (DD), despite the gate time being more
than one order of magnitude longer than the intrinsic coherence time
of the system. For realizing this universal conditional quantum gate,
we have devised a concatenated DD
sequence that ensures robustness against imperfections of DD pulses that otherwise may
destroy quantum information or interfere with gate dynamics. We compare its performance with
three other types of DD sequences. These experiments are carried out using a
well-controlled prototype quantum system -- trapped atomic ions
coupled by an effective spin-spin interaction. The scheme for
protecting conditional quantum gates demonstrated here is applicable
to other physical systems, such as nitrogen vacancy centers, solid
state nuclear magnetic resonance, and circuit quantum
electrodynamics.

\begin{description}\item[PACS numbers]
\verb03.67.Pp 03.67.Lx 03.65.Ud 37.10.Ty

\end{description}

\end{abstract}


\maketitle

Quantum information science has grown into an interdisciplinary
research field encompassing the investigation of fundamental
questions of quantum physics \cite{Zeilinger2010}, metrology
\cite{Schmidt2005,Roos2006} as well as the quest for a
quantum computer, or quantum simulator. The latter would allow
unprecedented insight into scientific problems relevant, for
instance, for physics and
chemistry \cite{Lloyd1996,Johanning2009b,Mueller2012,Jordan2012}. In order to exploit the principles of
quantum physics for such purposes, it is necessary to preserve
quantum coherence while carrying out gate operations. Dynamical
decoupling (DD) \cite{Viola1998,Khodjasteh2005} was successfully employed to extend the coherence
time of quantum states \cite{Biercuk2009,Du2009,deLange2010,Souza2011,Szwer2011,Wang2011,Gustavsson2012}, of single-qubit
operations \cite{Souza2012}, and of two-qubit quantum gates
\cite{vanderSar2012,Shulman2012} using pulsed fields. 
Also, dynamical control approaches that rely on shaped continuous fields have been suggested (e.g., \cite{Gordon2008,Schulte2005}) and implemented  (e.g., \cite{Timoney2008}).

The most eminent conditional quantum gate is the  two-qubit
Controlled-NOT (CNOT) gate, since it is a basic ingredient for
arbitrary quantum algorithms \cite{Barenco1995}. Physical systems
with which, in principle, such gates may be realized often do not
possess coherence times long enough compared to the time necessary
to carry out a gate. The coherence time may be limited by undesired
interactions of qubits with their environment and among themselves.
This reduces the achievable gate fidelity or  prevents conditional
quantum gates altogether.  It is, therefore, desirable to protect
the quantum system during its coherent evolution while carrying out
a conditional quantum gate.
One way to achieve this would be through the use of dynamical decoupling (DD) techniques \cite{West2010,Timoney2011}.

DD techniques, developed in the framework of nuclear magnetic
resonance (NMR), were originally intended to be used in high
precision magnetic spectroscopy \cite{Carr1954,Meiboom1958}.
In the modern field of quantum information processing, it was
investigated how the dephasing of a qubit, which would cause loss of
information during processing, could be suppressed by the use of DD
techniques
\cite{Viola1998}. New DD pulse sequences were proposed that are
optimal in particular environments or robust against operational
imperfections \cite{Khodjasteh2005}. The performance of DD sequences
in protecting the state of qubits (a quantum memory) was
successfully demonstrated, for example with ensembles of trapped
ions \cite{Biercuk2009}, individual ions \cite{Szwer2011}, solid
state NMR \cite{Du2009}, and quantum dots \cite{deLange2010}. In
addition to such single-qubit quantum memory investigations, 
experimental steps have been undertaken towards an entangled
two-qubit quantum memory, whose coherence time is enhanced by DD
pulses \cite{Wang2011,Gustavsson2012}.
A conditional gate protected by DD was demonstrated using a hybrid two-qubit systems with the qubits  dephasing at different time scales
\cite{vanderSar2012}.
A two-qubit gate with quantum dots was
performed using a single
spin echo pulse \cite{Shulman2012}, and a  two-qubit gate with trapped ions was made robust against variations of the driving fields' detuning using shaped pulses  \cite{Hayes2012}.

Our experiments are
carried out using two hyperfine qubits of trapped
atomic $^{171}$Yb$^+$-ions (a spin pseudo-molecule
\cite{Khromova2012}). 
The Hamiltonian describing this system reads
\begin{eqnarray}
    H=\frac{\hbar}{2} \omega_0^{(1)}\sigma_z^{(1)} +\frac{\hbar}{2} \omega_0^{(2)}\sigma_z^{(2)} -\frac{\hbar}{2}J_{12} \sigma_z^{(1)} \sigma_z^{(2)} \, ,
\end{eqnarray}
where $\omega_0^{(i)}$ is the resonance frequency of qubit $i$ and $\sigma_z^{(i)}$ is a Pauli matrix. We realize a physical system described by such a generic Hamiltonian
with two laser cooled $^{171}$Yb$^+$ ions, forming a pseudomolecule
\cite{Wunderlich2003,Johanning2009,Khromova2012} (see Supplemental
Material \cite{Supplement}). This Ising spin-spin coupling together
with single qubit rotations can be used to realize a conditional NOT
gate (CNOT) between a target and a control qubit
\cite{Supplement,Vandersypen2004}. This realization can be viewed as
a Ramsey-type experiment on the target qubit where first, a clockwise $\pi/2$ rotation around the x-axis of the Bloch sphere is applied and finally the same rotation around the y-axis (both driven by microwave radiation). 
During the conditional evolution time $T_g$ between these two pulses the target acquires a phase shift conditioned on the state of the control qubit. For the
experiments described below the $J$-coupling between control and
target qubit yields a necessary gate time of $T_g = 5$ ms.

\begin{figure}
\includegraphics[width=\columnwidth]{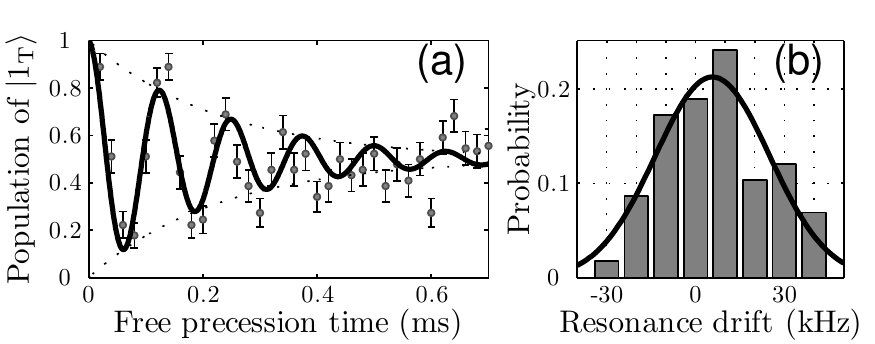}
\caption{ Dephasing and drift. (a) A Ramsey experiment is performed
to deduce the dephasing time of a qubit.  
A fit to the data yields a characteristic decay time of the fringes,
the dephasing time, of $200 \pm 100$ $\mu$s. We repeat the sequence
for every data point ($n=50$);error bars, s. d. (b) Drift between
consecutive experiments.
Before and after an experimental run that takes a few
minutes of data acquisition a qubit's addressing frequency is
measured by microwave-optical double resonance spectroscopy
The drift is Gaussian distributed with a width of 20 kHz, owing to
the stochastic origin of this process; $n=64$. }
\label{fig:RamseyDrift}
\end{figure}

When performing experiments fast magnetic field fluctuations are
present. We use DD pulses to probe the shape of the noise spectrum
\cite{Medford2012}, which yields a power-law $S(f) \propto f^{-2}$
in the range between 1 kHz and 50 kHz. These components, which are
much faster than the coupling, cause dephasing of quantum
superposition states within $200 \pm 100$ $\mu$s during the gate
operation (Fig. \ref{fig:RamseyDrift}(a)). This time scale, after
which we can not expect to observe any \textit{quantumness}, is more
than one order of magnitude shorter than the necessary gate time.
Therefore, it seems impossible to implement a quantum gate as the
one described above.

In addition to fast fluctuating magnetic fields, there are also
slowly varying stray fields present causing drifts of the qubits'
resonance frequency. A detuning of $\delta$ causes the spin vector
to rotate around an axis tilted by
$\arctan\left(\delta/\Omega\right)$ out of the x-y plane, where
$\Omega$ is the Rabi frequency of the qubit transition
\cite{Vandersypen2004}. The nutation angle of the pulse is also
relatively boosted by a factor of $\sqrt{1+(\delta/\Omega)^2}$. The
drift between consecutive experiments can be described by a Gaussian
distribution with a width of 20 kHz (fig. \ref{fig:RamseyDrift}(b)),
which is substantial in comparison with the Rabi frequency of about
$\Omega = 2\pi \times 60$ kHz. Therefore, the robustness of DD
techniques against instrumental errors is an important  feature to
be considered, not only in our trapped ion setup, but also, for
example, for nitrogen vacancy centers in diamond \cite{deLange2010} or other solid
state systems \cite{Souza2011}.

DD usually refocuses dephasing due to the qubits' interaction with
the environment but also couplings between qubits that are needed
for conditional quantum gates. We apply sequences of DD to both qubits 
simultaneously. The simultaneity is necessary, because this allows 
to enhance the coherence time of each individual qubit while not 
refocussing their common spin-spin interaction. DD as demonstrated 
here allows for the exchange of the roles of control and target qubit
with minimal modifications.

\begin{figure}
\includegraphics[width=\columnwidth]{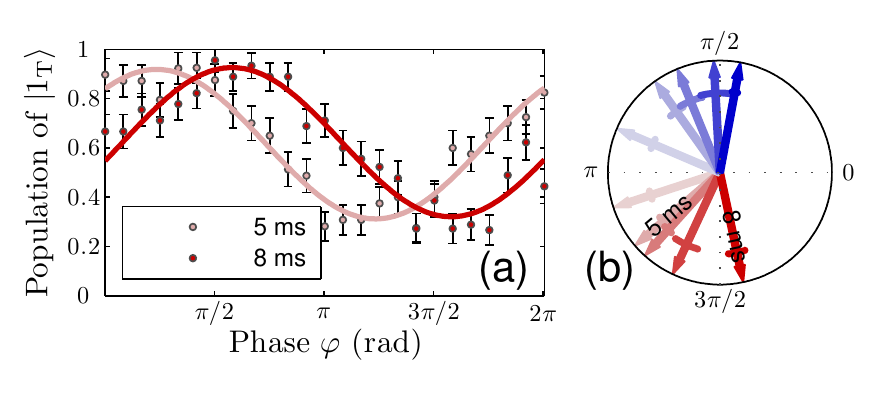}
\caption{ Conditional quantum dynamics. (a) 
A Ramsey-type experiment is
performed on the target qubit while DD pulses are applied to both
qubits simultaneously. 
The minimum of the Ramsey fringes, which is originally found
at $\pi$, is shifted when the conditional evolution time is
increased to 5 ms and 8 ms. $n=50$; error bars s. d. (b) Ramsey fringe
minima after 2 ms, 5 ms, 6 ms, 7 ms, and 8 ms of conditional evolution time.
After preparing the two-qubit system in $\ket{00}$ (red arrows), the
minima are shifted to higher Ramsey pulse phases $\varphi$. If the
control qubit is initialized in $\ket{1}$ (blue arrows), the target
precesses in the opposite direction; $n=10$; error bars s. d. }
\label{fig:FringeShift}
\end{figure}

To demonstrate conditional dynamics while DD pulses are applied, 
Ramsey interference experiments are carried out on the target qubit 
while the control is first prepared in $\ket{0}$. Between the two 
Ramsey pulses, during the time interval $T$ where the two qubits interact,  a DD sequence 
is applied. Ramsey fringes are observed by varying the phase of 
the second Ramsey pulse (figure \ref{fig:FringeShift}(a), here $T= 5$ ms or $T=8$ ms). 
The target qubits  phase 
coherence is now protected from dephasing during $T=8$ ms. This time has to be compared to the coherence time of 0.2 ms without DD (fig. \ref{fig:RamseyDrift}(a)). 

If the conditional 
evolution time $T$ between two Ramsey pulses is zero, the minima of 
the fringes are found at $\varphi=\pi$ (not shown in fig. \ref{fig:FringeShift}(a)). 
When the time $T$ is
increased, the Ramsey fringes are shifted accordingly, which reveals
precession of the target spin and is visualized in figure
\ref{fig:FringeShift}(b). The conditionality of these dynamics is
demonstrated by repeating the experiment with the control qubit
prepared in $\ket{1}$. If the conditional evolution time $T$ equals
the gate time $T_g$ (8 ms for the sequence used here; see below),
the minima are either to be found at $\varphi=3\pi/2$ or at
$\varphi=\pi/2$. Therefore, a Ramsey pulse with a chosen phase of
$\varphi=3\pi/2$ would leave the target in the state $\ket{0}$, or
flip it to the excited state $\ket{1}$ dependent on the state of the
control qubit, thus realizing a conditional spin flip. Discussion of gate 
fidelities deduced from Ramsey fringe contrast and additional measurements can be found in \cite{Supplement}.

\begin{figure}
\includegraphics[width=0.8\columnwidth]{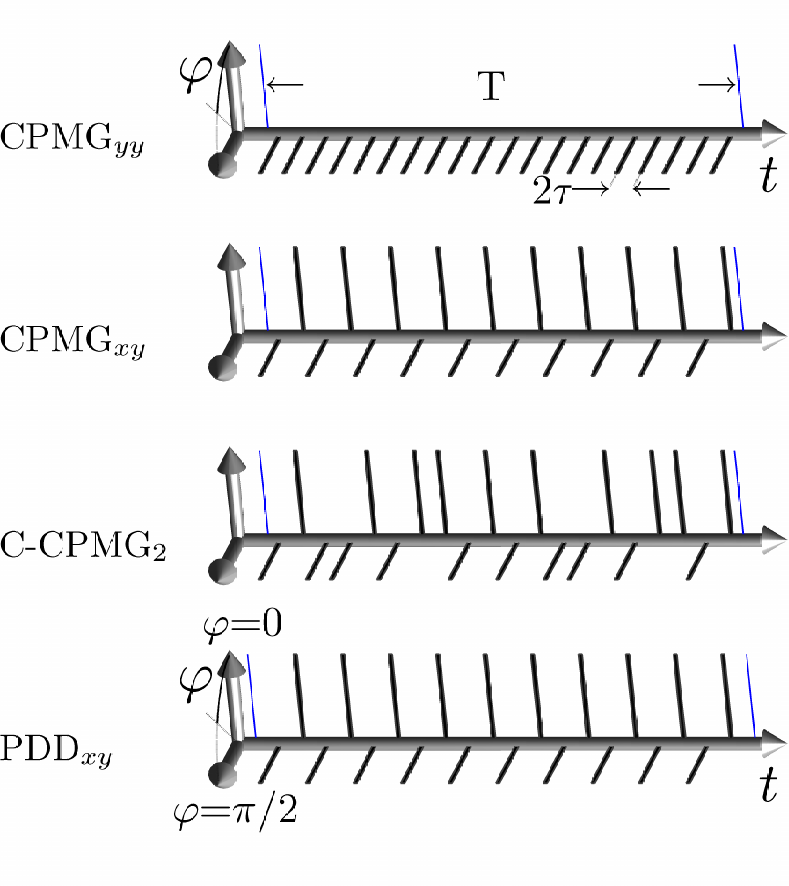}
\caption{Four DD sequences employed to protect conditional quantum
gate dynamics. $\pi$ pulses (black lines) are applied on both qubits
during a conditional evolution time $T$ between two Ramsey $\pi/2$
pulses (blue lines) applied only to the target qubit. All the
sequences show the same pulse interval, except the PDD sequence for
the first and last $\pi/2$ pulses. To achieve robust sequences the
pulses are applied with different phases ($\varphi=0$ and
$\varphi=\pi/2$).} \label{fig:Sequences}
\end{figure}

In what follows, we first describe four different DD sequences (fig.
\ref{fig:Sequences}) that are investigated experimentally, and
thereafter discuss their performance in protecting the desired
quantum gate dynamics from dephasing. The first sequence is labelled
$\text{CPMG}_\text{yy}$ since it is a Carr-Purcell-Meiboom-Gill
sequence \cite{Carr1954,Meiboom1958} that consists of $\pi$ pulses
rotating the qubit  around the y axis only. This sequence protects a quantum
state whose Bloch vector lies along the y axis. When considering the
described gate, where the target spin precesses, we can not expect
this sequence to perform well. To protect any arbitrary
superposition state, one should make use of isotropic
\cite{Souza2011} sequences that rotate around x and y axes equally.
This can be accomplished by alternating the pulse phases between
$\varphi=0$ and $\pi$. It is possible to further improve the
robustness of a sequence by the use of concatenated sequences
\cite{Souza2011}. In these sequences individual pulse errors do not
accumulate but compensate each other.

We introduce a new kind of concatenated DD sequence, where the phases of the individual pulses are constructed by a concatenation, while their timing still follows CPMG-type sequences. 
As the basic sequence we define $\text{C-CPMG}_\text{1}\equiv [\tau Y \tau
\tau \bar{X} \tau]^2$, where $Y$ and $\bar{X}$ denote $\pi$
pulses rotating the qubit anti-clockwise around the y or clockwise around the x axis and $[]^2$ indicates a repetition of the time evolution in parentheses. 
The next levels of our concatenated CPMG sequence are constructed by the recursion formula $\text{C-CPMG}_\text{n}=[\sqrt{\text{C-CPMG}_\text{(n-1)}} \tau Y
\tau \text{C-CPMG}_\text{(n-1) } \tau \bar{X} \tau
\sqrt{\text{C-CPMG}_\text{(n-1)}}]^2$, $n\geq2$. 
In contrast to the original concatenated sequences \cite{Souza2011} the additional conditional evolution times $\tau$ around the pulses still feature the CPMG timing.

Another sequence investigated is an isotropic, strictly periodic
$\text{PDD}_{xy}$ sequence \cite{Viola1998}. Its timing deviates
from the CPMG sequence only in the fact that the first and last
intervals are of the same length as any other one. At first sight
this may seem to be only a slight change, but in fact substantially
changes the performance of the pulse sequence. This can be
understood by considering that a DD sequence can be viewed as a
filter function cutting out particular parts of a noise
spectrum. Hence, different sequences are described by different
filter functions and thus have a different performance under
particular noise conditions \cite{Biercuk2009}.

We compare the performance of the DD sequences described above by
Ramsey interference experiments that are carried out after preparing
the system in each of the four input states of a CNOT truth table
($\ket{00}, \ket{01}, \ket{10}$ and $\ket{11}$). When using a
$\text{CPMG}_{yy}$ sequence of 24 DD pulses during 5 ms of
conditional evolution time, we can observe Ramsey fringes with a
clear contrast indicating preserved single qubit phase coherence. 
This demonstrates the capability of protecting the
quantum system from its noisy environment. However, we cannot
observe any significant conditional phase shift. This behaviour is
reproduced by simulating the experiment while taking the detunings,
originating from the drift (fig. \ref{fig:RamseyDrift}(b)),
explicitly into account. The results of our simulation indicate that
the pulse imperfections in the $\text{CPMG}_{yy}$ sequence suppress
the precession and thus hamper the desired evolution, which renders the conditional gate realization impossible.

\begin{figure}
\includegraphics[width=\columnwidth]{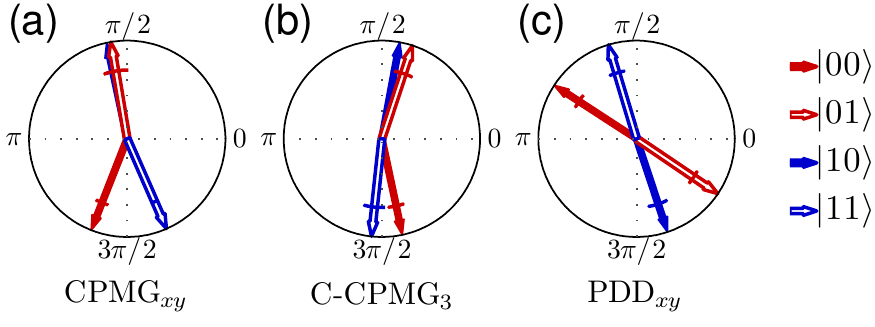}
\caption{ Experimental conditional gate evolution under three
different DD sequences. 
Ramsey fringes are recorded and
the phase of their minima is plotted. If we apply $\text{CPMG}_{xy}$
(a, 24 pulses)  or $\text{C-CPMG}_\text{3}$ (b, 84 pulses) during
conditional evolution time, the minima are either found around
$\varphi=3\pi/2$ or $\varphi=\pi/2$ depending on the state of the
control qubit. If, however a $\text{PDD}_{xy}$ sequence is used the
minima gain an additional global shift of 0.8 rad (see text for
details). The relative phase shift between different input states of
the control is smaller than $\pi$. Data are an average of about 10
Ramsey type experiments (25 phase-points and $n=100$ repetitions);
error bars, s. d. } \label{fig:FringeGates}
\end{figure}

Figure \ref{fig:FringeGates} presents the results for the three
other sequences. When using the $\text{CPMG}_{xy}$ sequence with 24
alternating pulses or  $\text{C-CPMG}_{3}$ with 84 pulses, the clearly
observed Ramsey fringes show the conditional phase shifts as
expected (\ref{fig:FringeGates}(a) and (b)). A Ramsey pulse of phase
$\varphi=3\pi/2$ flips the target qubit only if the control is in
$\ket{1}$, which defines the CNOT gate. Both sequences give similar
results in terms of the conditional phase shift and the gate
fidelity \cite{Supplement} when using a $J$-coupling that result in
a gate time $T_g=5$ ms. However, for longer gate times, the
performance of $\text{CPMG}_{xy}$ sequence degrades rapidly while
$\text{C-CPMG}_{3}$ is able to protect longer quantum gates as well \cite{Supplement}.  
For the concatenated sequence the time between
the two Ramsey pulses is increased from 5 to 8 ms. We explain this
by two different effects. First, the duty cycle of 84 pulses reduces
the effective time in which the spin may conditionally precess, and
second, imperfections of the larger number of pulses still results
in a slightly suppressed precession.

For the strictly periodic sequence $\text{PDD}_{xy}$, 49 pulses yield
the best fringe contrast, but the relative phase shift resulting
from different control qubit input states is smaller than $\pi$.
Here, also the duty cycle of the pulses and imperfections impair the
spin precession. However, it is not possible to further increase the
conditional evolution time because of the sequence's low capability
to prevent the spin from dephasing. Increasing the number of pulses
does not improve the results as now pulse imperfections
strongly harm the dynamics as reflected by a reduced fringe contrast. In addition, the fringes are no longer
centred around $\pi$ but instead, the target qubit is additionally
rotated 0.8 radians around the z axis. We explain this spurious
extra phase shift by an oscillating magnetic field along the
quantization axis. If DD pulses are applied always when the
oscillation changes its sign, the pulses will not cancel the
acquired precession angles but cause them to add up. This coherent
phase pickup may be exploited to realize a single-ion quantum
lock-in amplifier \cite{Kotler2011}. We identified a dominant
noise source in the range of a few kHz that is correlated to DD sequences. This noise is picked up by the power supply and the cables connecting to the Helmholtz coils that define the quantization axis. This in turn leads to magnetic field oscillations correlated with the pulse sequence. For the other DD sequences this effect was not observed, because of the different delay between the pulses. Beside this additional shift, also
the odd number of DD pulses has to be considered. This results in an
additional rotation that shifts the position of all minima by $\pi$.

\begin{figure}
\includegraphics[width=\columnwidth]{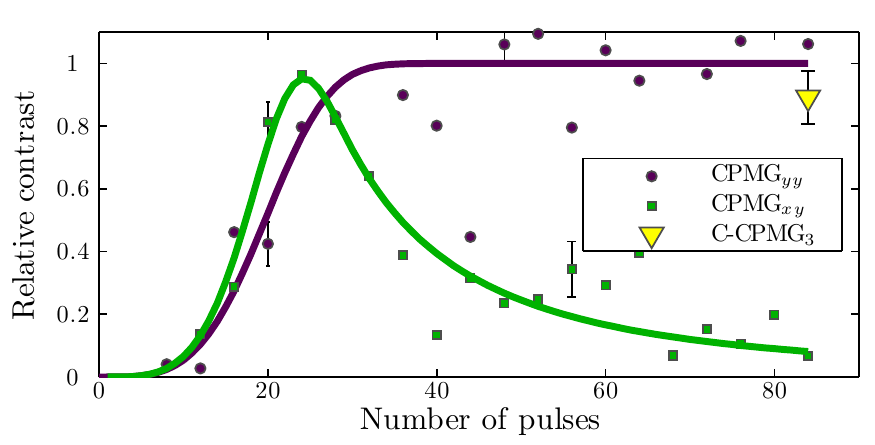}
\caption{ Robustness of the DD sequences. On one of the qubits a
Ramsey experiment is performed. During a free evolution time of 5 ms
DD sequences are applied and the final state is probed. The phase of
the final Ramsey pulse is chosen as $\varphi=0$ in order to detect
the fringe maximum. $\text{CPMG}_{yy}$ is used as a benchmark
(however, does not permit a conditional quantum gate). The
$\text{C-CPMG}_\text{3}$ sequence performs well even when many
pulses are used, and, thus makes conditional quantum gates possible
for gate times long compared with the coherence time. Data are
averaged over ten experimental runs with 100 repetitions of each
data point. Lines are drawn to guide the eye; error bars s. d. }
\label{fig:DutyCycle}
\end{figure}

So far, we have investigated the performance of different sequences
to protect the conditional precession of a dephasing spin while
being coupled to a control qubit. Now, we investigate in more detail
how imperfections in successfully tested sequences may impair the
gate dynamics. For this purpose, again a Ramsey experiment is
carried out on the target qubit. In contrast to the experiments
before, DD pulses are addressed to the target only which effectively
refocuses the spin-spin coupling \cite{Vandersypen2004} and prevents
any conditional precession. The fringe contrast as a function of the
number of DD pulses is shown in figure \ref{fig:DutyCycle}. The
$\text{CPMG}_{yy}$ is robust for the particular input state whose
Bloch vector lies along the y axis \cite{Souza2011}, and therefore
the contrast grows with the number of applied DD pulses. For 24
pulses the coherence time is $10 \pm 1$ ms.  Then the contrast reaches a
plateau that can be considered as a benchmark. Using the
$\text{CPMG}_{xy}$ sequence, in contrast, pulse imperfections
accumulate and cause a seemingly chaotic behaviour for increasing
number of pulses: If more than 24 pulses are used, this random
behaviour reduces the contrast of Ramsey fringes after several
realizations of the experiment. The concatenated sequence
$\text{C-CPMG}_\text{3}$ with 84 pulses, when performed under the
same experimental conditions, still shows a high relative contrast.
This clearly demonstrates the fact that individual pulse
imperfections do not accumulate but compensate each other in this
sequence. Since a higher number of dynamical decoupling pulses also
yields a longer coherence time this sequence is also able to protect
slower gates, based on a lower $J$ coupling strength.
For comparison, by the use of a $\text{CPMG}_{xy}$ sequence, realized by simply adding more and more alternating pulses, it is not possible to implement a slower gate.

Due to its generality, the approach demonstrated here for carrying out quantum gates in a environment that causes dephasing is applicable to a large variety of physical systems.
Thus, existing quantum gates could be improved
to reach the fidelity limit that would allow for scalable fault-tolerant quantum computing.
The universal idea of protecting coherent quantum dynamics could be applied as well in other contexts where conditional quantum logic is used, for example, for a quantum repeater,  metrology, or spectroscopic applications.

Technical help with the microwave
set-up by T. F. Gloger is acknowledged. We thank M. Johanning for
discussions. We acknowledge funding by the Bundesministerium
f\"{u}r Bildung und Forschung (FK 01BQ1012), Deutsche
Forschungsgemeinschaft, and from the
European Community's Seventh Framework Programme (FP7/2007-2013) under
grant agreement number 270843 (iQIT) and number 249958 (PICC).

\end{document}